\documentclass[%
 reprint,
 amsmath,amssymb,
pra,
]{revtex4-2}

\usepackage{graphicx}
\usepackage{dcolumn}
\usepackage{bm}
\usepackage[version=4]{mhchem}
\usepackage{hyperref}

\newcommand{\Zeff}{Z_\text{eff}}
\newcommand{\EB}{\varepsilon_{b}} 	
\newcommand{\Npi}{N_{\pi}}

\usepackage{titlesec}
\titlespacing*{\subsubsection}{0.5pt}{\baselineskip}{0pt}

\usepackage{soul}
\usepackage{xcolor}

\begin{document}

\title{Positron annihilation and binding in aromatic and other ring molecules}

\author{E. Arthur-Baidoo}
 \altaffiliation{}
 \email{earthurbaidoo@ucsd.edu}
 \author{J. R. Danielson}%
 \email{jdan@physics.ucsd.edu}
 \altaffiliation{}
 \author{C. M. Surko}%
 \altaffiliation{}
 \email{csurko@physics.ucsd.edu}
 \affiliation{Department of Physics, University of California San Diego, La Jolla, CA, 92093 USA.}
 
 \author{J. P. Cassidy$^1$}%
 \affiliation{{$^1$}Centre for Light-Matter Interactions, School of Mathematics and Physics,
Queen's University Belfast, \\University Road, Belfast BT7 1NN, Northern Ireland, United Kingdom.\\ {$^2$}School of Physics, Trinity College Dublin, Dublin 2, Ireland.}%
 \author{S. K. Gregg$^1$}%
 \affiliation{{$^1$}Centre for Light-Matter Interactions, School of Mathematics and Physics,
Queen's University Belfast, \\University Road, Belfast BT7 1NN, Northern Ireland, United Kingdom.\\ {$^2$}School of Physics, Trinity College Dublin, Dublin 2, Ireland.}%
 \author{J. Hofierka$^1$}%
 \affiliation{{$^1$}Centre for Light-Matter Interactions, School of Mathematics and Physics,
Queen's University Belfast, \\University Road, Belfast BT7 1NN, Northern Ireland, United Kingdom.\\ {$^2$}School of Physics, Trinity College Dublin, Dublin 2, Ireland.}%
 \author{B. Cunningham$^1$}%
 \affiliation{{$^1$}Centre for Light-Matter Interactions, School of Mathematics and Physics,
Queen's University Belfast, \\University Road, Belfast BT7 1NN, Northern Ireland, United Kingdom.\\ {$^2$}School of Physics, Trinity College Dublin, Dublin 2, Ireland.}%
 \author{C. H. Patterson$^2$}%
 \author{D. G. Green$^1$}%
 \email{d.green@qub.ac.uk}
 \affiliation{{$^1$}Centre for Light-Matter Interactions, School of Mathematics and Physics,
Queen's University Belfast, \\University Road, Belfast BT7 1NN, Northern Ireland, United Kingdom.\\ {$^2$}School of Physics, Trinity College Dublin, Dublin 2, Ireland.}%

\date{\today}

\begin{abstract}
Annihilation spectra are presented for aromatic and heterocyclic ring molecules resolved as a function of incident positron energy using a trap-based positron beam. Comparisons with the vibrational mode spectra yield positron-molecule binding energies. Good to excellent agreement is found between the measured binding energies and the predictions of an \textit{ab initio} many-body theory that takes proper account of electron-positron correlations including virtual-positronium formation. The calculations elucidate the competition between permanent dipole moments and $\pi$ bonds in determining the spatial distribution of the bound-state positron density. The implications of these results and the role of multimode features in annihilation in these molecules, including Fermi resonances, are discussed.
\end{abstract}

\maketitle

\section{\label{sec:level1}INTRODUCTION AND OVERVIEW}

Positrons bind to many polyatomic molecules via the excitation of vibrational Feshbach resonances (VFR) \cite{Surko88,Grib10,Grib09}. Binding energies for over 100 molecules have been measured to date with magnitudes ranging from a few millielectron volts to $>$ 0.3 eV. 
While positrons are also predicted to bind to atoms \cite{PhysRevA.52.4541,PhysRevLett.79.4124,Strasburger:1998},
they have yet to be observed experimentally due to the lack of low-lying excitations to mediate attachment, though proposals exist \cite{Mitroy1999,PhysRevLett.105.203401,Surko2012,Swann2016}. Thus the molecular studies play a special role in our understanding of positron interactions with matter.

Experimentally, positron binding has been observed to depend upon molecular parameters such as the polarizability $\alpha$, permanent dipole moment $\mu$, the number of molecular $\pi$ bonds {$N_\pi$}, the ionization potential $I$, and the geometry (e.g., upon isomerization) \cite{Danielson09,Danielson23,Swann21,danielson2012interplay,Dan22}.
Theoretical work successfully predicts many aspects of these chemical trends, but there remain a number of important questions \cite{Sugiura19,Tachikawa14,Tachikawa20,Swann19,Swann21,Suguira20,cassidy2023manybody}.
For unsaturated ring molecules, enhancement of the positron-molecule binding energy $\varepsilon_{\textit{\text{b}}}$  with increasing $\Npi$ is observed \cite{Dan22}.

A recent many-body theory (MBT) \cite{Hofierka22} developed by some of us  
provides an \textit{ab initio} description of positron binding to molecules, importantly taking proper account of electron-positron correlations including the nonperturbative process of virtual-positronium formation (where a molecular electron temporarily tunnels to the positron). To date it has, e.g., provided the first calculated binding energies in agreement with measurements, quantified the role of correlations and the contributions of individual molecular orbitals to the positron-molecule correlation potential, and explained trends within molecular families \cite{Hofierka22, cassidy2023manybody}. It has recently been extended to positron scattering and annihilation on small molecules \cite{Rawlins:2023}. Particularly relevant to the present study, the MBT provides new insights through calculations of the spatial distribution of the bound-state positron wave functions.

In this paper, measurement of the positron annihilation spectrum and binding energy are presented for aromatic and substituted ring molecules and compared to predictions of the many-body theory. 
This work extends earlier studies \cite{Dan22,danielson2012interplay,Danielson09} using ring substitutions to study molecules with varying amplitudes of permanent dipole moment and numbers of $\pi$ bonds. As a result, it provides a sensitive test of the competition between these two mechanisms of positron binding.

Results are presented for benzene, toluene, acetophenone, aniline, phenylacetylene, benzonitrile, benzaldehyde, furan, pyrrole, pyridine, pyridazine, cycloheptatriene, and cyclooctatetraene. The molecules studied span a wide range of {$N_\pi$} (2--5) and $\mu$ (0.3--4.5 D) and a factor of 1.7 in $\alpha$. Good to excellent agreement is found between the MBT predictions and binding energy measurements for this new class of molecules, which include the largest studied to date via the MBT.

As discussed below, the MBT shows that regions of the bound-state positron density are localized near the $\pi$-bond $\emph{electron density}$ above and below planes of ring molecules such as benzene and toluene. This exemplifies the importance of the $\pi$-bonds in localizing the bound-state positron wave function. 
Other distributions of positron density are localized close to strong permanent dipole moments that typically lie in the planes of ring molecules such as in benzaldehyde and pyridine. For molecules with weaker dipole moments, including furan and pyrrole, the effects compete, and the bound-state positron distribution is observed in regions of both the $\pi$-bond electrons and adjacent the dipole moment. This contrasts to the approximately uniform covering of positron density surrounding non-polar molecules without $\pi$ bonds such as that seen in alkanes \cite{Grib09}.

Positron annihilation rates for molecules are conventionally expressed in terms of the quantity $Z_\text{eff}$ which is the measured rate normalized by that expected for a gas of free electrons with the density equal to the density of the molecular gas \cite{Grib10}. Positron attachment via VFR typically results in annihilation rates $Z_\text{eff} \gg 1$ and greater than those expected for a simple collision (e.g., $Z_\text{eff} \simeq$ the number of valence electrons) \cite{Grib10}. The VFR theory of positron annihilation due to dipole- and quadrupole-allowed fundamental vibrations can account for enhancements of an order of magnitude or so in annihilation rate (e.g., $Z_\text{eff} \le 2000$ for a single vibrational mode) \cite{Grib06,Natisin17}. However, $Z_\text{eff}$ for many molecules, including those studied here, is 2 -- 3 orders of magnitude larger, and this is not understood. 

Of note in this study, and similar to that observed previously in benzene \cite{ghosh2022resonant}, broad regions in the annihilation spectra 
are observed in aromatic molecules such as pyridine and aniline at incident positron energies that do not correspond to fundamental vibrations. The origins of these features are likely multimode vibrations where the specific modes involved have yet to be identified. Related to this, enhanced annihilation is observed in benzaldehyde that is likely due to the multimode phenomenon of Fermi resonance (the resonance of a fundamental vibration with the second harmonic of another vibration). This provides an example of a multi-mode contribution to the annihilation spectrum where the specific modes involved are known and is worthy of further study.

This paper is organized as follows. Section II presents the description of the experiment and data analysis procedures. Section III presents the measured annihilation spectra as a function of incident positron energy and the resulting binding energy analyses. Section IV describes the many-body theory and the predicted binding energies for the molecules studied here and comparison with the measurements. Further aspects of the results are discussed in Sec.~V, and Sec.~VI presents a summary and concluding remarks.

\section{\label{sec:level1}DESCRIPTION OF THE EXPERIMENT AND DATA ANALYSIS}
The experimental techniques used in this positron annihilation study have been described in detail previously  \cite{Danielson23}. Low-energy positrons are obtained from a $^{22}$Na radioisotope source and a solid neon moderator. Positrons leaving the moderator are radially confined and guided using  magnetic fields into a three-stage buffer-gas trap (BGT) \cite{Nat15}. In the BGT, they are slowed by inelastic collisions with N$_2$ and CF$_4$ resulting in positrons trapped and cooled to the ambient temperature (300 K). 
The pulsed positron beam is formed by slowly ramping the bottom of the potential well to a voltage higher than that of the exit gate \cite{Nat15,Danielson23}. The beam thus formed has a narrow energy spread with a mean energy just slightly larger than the exit gate potential. The energy distribution of the beam is measured using a retarding potential analyzer (RPA). 

For these experiments, the \emph{parallel} energy distribution of the beam is approximately Gaussian with a mean energy $E_{\parallel} =$ 0.67--0.70 eV and standard deviation $\sigma_{\parallel}$= 8--10 meV. The perpendicular energy of the beam is obtained from measuring mean parallel energy at different RPA magnetic fields. The slope of this measurement yields the mean perpendicular energy \textit{E$_\perp$}= 20 $\pm$ 2 meV \cite{Nat15}. The distribution in total energies is the convolution of the parallel and perpendicular energy distributions which yields an exponentially modified Gaussian distribution (EMG) \cite{Nat15}.
The guiding magnetic field varies by a factor of five along the beam line, where the annihilation gas cell is located at the end \cite{ghosh2020energy}. The magnetic field at the gas cell is independently controlled. When it matches the value at the exit gate of the BGT, both the parallel and perpendicular particle distributions are known, and hence the total energy distribution is known for positrons interacting with the test gas.

The positrons propagate downstream to a 26 cm long gas cell, the potential of which is used to set the mean parallel energy of the beam interacting with the test gas. During annihilation measurements, an isolated electrode after the annihilation cell is biased to 6 V to reflect the beam back towards the BGT. Positrons continue to bounce between the exit gate of the BGT and the reflecting electrode while measurements are made. See Ref. \cite{Danielson23} for details.

The measured annihilation rates are obtained by converting the annihilation counts (\textit{$N_c$}) to an annihilation cross section using the number of positrons per pulse (\textit{$N_p$}), the calibrated detector efficiency ($\eta_{D}$), the test-gas density ($n_{m}$) and the annihilation cell length ($L_{D}$). For a given molecule, the normalized annihilation rate is given by \cite{Grib10}
\begin{equation}
Z_{\text{eff}}(E_{\parallel}) = \frac{N_{\text{c}}(E_{\parallel})}{2N_{\text{p}}\eta_{\text{D}}L_{\text{D}}n_{\text{m}}}\frac{\upsilon(E_{\parallel})}{\pi r_{0}^{2}c}
\end{equation}
where $\upsilon(E_{\parallel})$ is the mean particle velocity, $r_0$ is the classical radius of the electron, and \textit{c} is the speed of light. The factor of two in the denominator accounts for the two passes of the positrons through the annihilation cell in a single bounce.

Shown in Fig.~1 are contemporary data for the $Z_\text{eff}$ of the alkane octane plotted as a function of the mean parallel energy of the beam. 
Also shown is the downshifted IR spectrum (arbitrarily scaled) \cite{linstorm1998nist}, which demonstrates the spread of the vibrational modes.
The $\Zeff$ spectrum shows an asymmetric resonant peak near 200 meV. 
In order to obtain the binding energy, a fit to the C-H stretch resonance is performed where each IR-active mode is taken to be a delta function and convolved with the beam EMG distribution. The resonances from all modes are summed to give a single resonant peak with a width which is a combination of the measured beam parameters and energies of the vibrational modes. The positron-molecule binding energy $\varepsilon_{\textit{\text{b}}}$ is obtained using the total energy $\varepsilon_r$ at which the resonance is observed and the vibrational mode energy $\hbar\omega$ with the relationship,
\begin{equation}
\varepsilon_{\textit{\text{b}}} = \hbar\omega - \varepsilon_r.
\end{equation}
For the octane data in Fig.~1, this procedure yields a binding energy of 150 $\pm$ 3 meV \cite{Danielson23}. 

\begin{figure}[tb]
\includegraphics[width=0.9\linewidth,keepaspectratio]{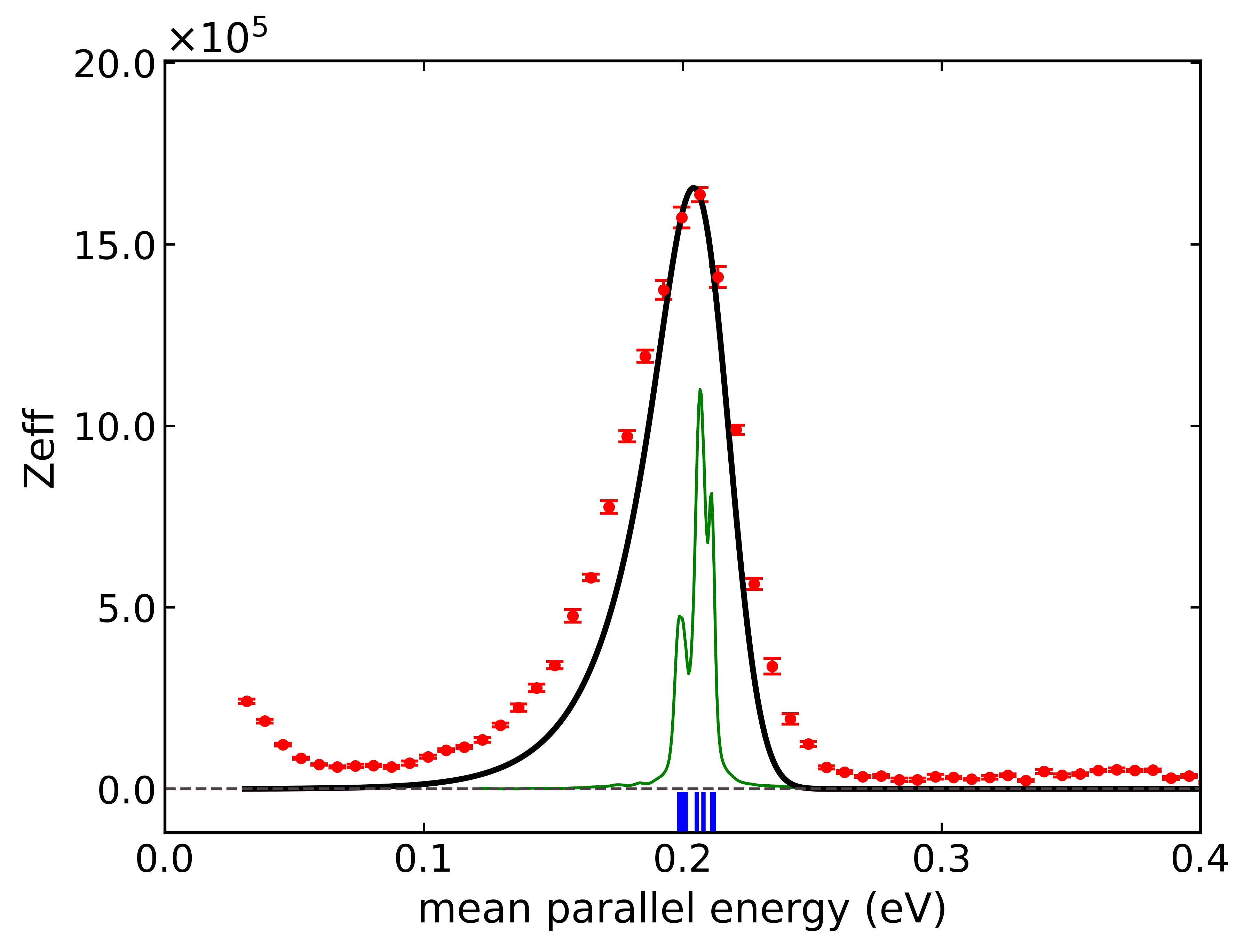}
\caption{\label{fig:epsart} Annihilation rate $Z_{\text{eff}}$ for octane as a function of mean parallel energy of the positron beam. The green curve is the IR spectrum of octane obtained from NIST \cite{linstorm1998nist}, and vertical blue bars indicate the locations of the IR active fundamental modes, with both downshifted by $\varepsilon_{b}$. The solid black line shows the fit to the C-H modes used to determine $\varepsilon_{b}$.}
\end{figure}

In contrast to octane and other alkane molecules \cite{Danielson23}, benzene and the other aromatics studied here have appreciable annihilation amplitudes that are believed to be due to unresolved non-fundamental modes \cite{ghosh2022resonant}.  This leads to uncertainty in obtaining the resonances used in Eq.~2. As a result, for all aromatic molecules, the choice was made to use only the dipole-active high-energy phenyl-CH stretch modes to obtain $\varepsilon_{\textit{\text{b}}}$ values. These modes used are shown as solid vertical lines in the figures, and the resulting fits are shown by the solid black lines. For most molecules, these fits match the data reasonably well. However, in a number of cases, including the other components of the spectra can modify the best-fit energy of the resonance by several meV. Thus, the conservative error estimate of $\pm$ 5 meV is associated with the $\varepsilon_{\textit{\text{b}}}$ measurements unless otherwise noted.

This analysis is based on the Gribakin and Lee (GL) theory for $Z_\text{eff}$ from VFRs mediated by dipole-allowed fundamental vibrational modes \cite{Grib06,Grib10}. If the IR activity for the fundamentals is known, and using $\EB$ obtained from above, the contribution from each mode can be summed to predict the total expected $Z_\text{eff}$ as a function of the mean parallel energy of the positron beam. For almost all molecular targets this estimate is \emph{much smaller} than that observed. However, the spectral shape is often quite accurate. 

The amplitude of the fit can be related to the enhancement factor that the GL prediction must be multiplied by to match the measurements. For the molecules here, this factor varies from $\sim 220$ for cyclooctatetraene to $\sim 1.5$ for furan. The solid line in Fig.~1 is 75 times the GL prediction for octane. This enhancement is believed to be due to the coupling of the fundamentals to a large number of multi-mode resonances by intramolecular vibrational redistribution (IVR) \cite{Stewart83,Grib10}. However, this has yet to be confirmed. The origin of this enhancement mechanism remains as a major topic of study, beyond the scope of the work presented here.

\section{\label{sec:level1}ANNIHILATION SPECTRA AND BINDING-ENERGY ANALYSES}
The current study focuses on the manner in which additions to, and substitutions in the benzene ring affect binding energies. The molecules studied can be grouped in three categories: (i) benzene (\ce{C6H6}) and its derivatives, toluene (\ce{-CH3}), acetophenone (\ce{-COCH3}), phenylacetylene (-CCH), benzonitrile (-CN),  and benzaldehyde (-CHO); (ii) the heterocyclic aromatic molecules furan, pyrrole, pyridine, and pyridazine; and (iii) non-aromatic cyclic molecules, cycloheptatriene (CHT) and cyclooctatetraene (COT). The results for each category and molecule are discussed separately. For easy referral during this discussion, the molecular parameters for all molecules and the measured positron-molecule binding energies are summarized in Table I. 

\begin{table*}
\caption{\label{tab:table3}Experimental (exp.) binding energies for the molecules studied, together with their molecular properties, polarizability ($\alpha$) \cite{haynes2014crc,johnson2013nist}, dipole moment ($\mu$) \cite{haynes2014crc,johnson2013nist}, number of pi-bonds ($N_{\pi}$), and ionization potential ($I$) \cite{haynes2014crc,johnson2013nist}. The corresponding many-body theory calculated binding energies from the present work are shown in Table \ref{tab:bind}}
\begin{ruledtabular}
\begin{tabular}{ccccccc}
 Molecule & Structure & N{$_\pi$} & $\mu$ [D] & $\alpha$ [$\text{\AA}^3$]& $I$ [eV] & $\varepsilon_{b}$ (exp.) [meV] \\
 \hline
 \multicolumn{7}{l}{\textbf{\textit{}}}\\
 \multicolumn{7}{l}{\textbf{\textit{Benzene and its substituted derivatives}}}\\ 
 \hline
  Benzene & \ce{C6H6} & 3 & 0.00 & 10  & 9.24 & 132$\pm$3 \footnote{Value taken from Ref. \cite{ghosh2022resonant}.} \\
 Toluene & \ce{C6H5CH3} & 3 & 0.38& 11.86  & 8.83 & 173$\pm$5 \\
 Acetophenone & \ce{C6H5COCH3} & 4 & 3.02 & 14.4   &9.28 & 288$\pm$5\\
 Aniline & \ce{C6H5NH2}  & 3 & 1.53 & 12.1   & 7.72 & 233$\pm$5\\
 Phenylacetylene & \ce{C6H5CCH} & 5 & 0.66 & 13.8  & 8.82 & 230$\pm$5\\
 Benzonitrile & \ce{C6H5CN} & 5 & 4.52 & 12.5  & 9.73 &298$\pm$5\\
 Benzaldehyde & \ce{C6H5CHO} & 4 & 3.14 & 12.80   & 9.50 & 220$\pm$10\\
 \hline
 \multicolumn{7}{l}{\textbf{\textit{}}}\\
 \multicolumn{7}{l}{\textbf{\textit{Heterocyclic aromatic molecules}}}\\ 
 \hline
 Furan & \ce{C4H4O} & 2 & 0.66 & 7.23  & 8.88 & 52$\pm$5 \\
 Pyrrole & \ce{C4H4NH} & 2 & 1.77 & 7.9    & 8.2 & 165$\pm$10 \\
 Pyridine & \ce{C5H5N} & 3 & 2.19 & 9.5   & 9.26 & 186$\pm$5\\
 Pyridazine & \ce{C4H4N2} & 3 & 4.22 & 9.27    & 8.74 & 330$\pm$5\\
 \hline
 \multicolumn{7}{l}{\textbf{\textit{}}}\\
 \multicolumn{7}{l}{\textbf{\textit{Non-aromatic cyclic molecules}}}\\ 
 \hline
 Cycloheptane & \ce{C7H14} & 0 & 0.00 & 12.8    & 9.97 & 104$\pm$4 \footnote{From Ref. \cite{Danielson23}} \label{footnote 1}\\
 Cycloheptatriene & \ce{C7H8} & 3 & 0.25 & 12.6    & 8.29 & 190$\pm$8\\
 Cyclooctane & \ce{C8H16} & 0 & 0.00 & 14.5    & 9.75 & 128$\pm$4 \textsuperscript{b}\\
 Cyclooctatetraene & \ce{C8H8} & 4 & 0.00 & 13.76    & 8.43 & 225$\pm$5
\end{tabular}
\end{ruledtabular}
\end{table*}

\subsection{Benzene and its substituted derivatives}
\begin{figure*}
\includegraphics[width=0.9\linewidth, keepaspectratio]{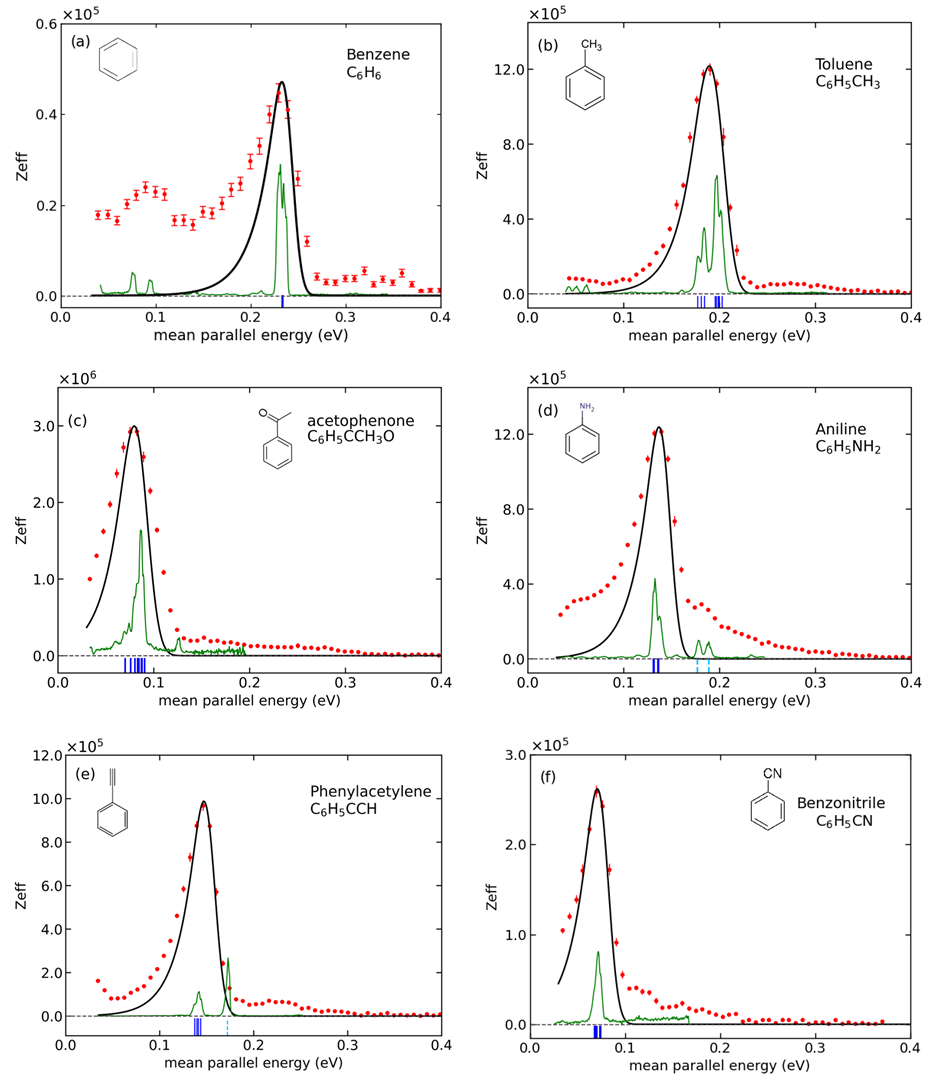}
\caption{\label{fig:wide1} Measured $Z_{\text{eff}}$ values for benzene and its substituted derivatives (a) benzene, (b) toluene, (c) acetophenone, (d) aniline, (e) phenylacetylene, (f) benzonitrile. The green curves are the IR spectra obtained from NIST \cite{linstorm1998nist}, and vertical blue bars indicate the locations of the IR active fundamental modes, with both downshifted by $\varepsilon_{b}$. The solid black curves are fits to the C-H peaks. Vertical dashed green lines indicate dipole active fundamental modes not used in the analysis.}
\end{figure*}

A high-resolution annihilation spectrum for benzene was measured previously using the cryogenic trap-based beam, yielding $\varepsilon_{b}$ = 132 $\pm$ 3 meV \cite{ghosh2022resonant}. However, to better compare to the other molecules, data from the room-temperature trap-based beam is shown in Fig.~\ref{fig:wide1}\,(a), along with the fit to the single dipole-allowed C-H vibrational mode.
As discussed in Ref.~\cite{ghosh2022resonant}, an unusual feature of the spectrum is a broad range (i.e., $\sim$ 0.3 - 1.5 eV) of enhanced annihilation in a region of energies absent fundamental vibrational modes.  As discussed below, these features appear to be quite common in aromatic molecules.

Benzene derivative molecules with side-group substitutions typically have dipole moments due to the addition or substitution. The new molecule with the smallest dipole moment studied here is toluene ($\mu$ = 0.33 D), which consists of a benzene ring with an attached methyl (CH$_3$) group. It has three $\pi$ bonds in the ring. While the net dipole moment of benzene is zero due to symmetry, the methyl group both creates a dipole moment and increases the polarizability of the molecule. 

Shown in Fig.~\ref{fig:wide1}\,(b) is the annihilation spectrum of toluene.  It has a sharp peak due to the five IR-active C-H stretch modes from benzene and three others from the methyl group that are visible in the IR spectrum \cite{fuson1960spectre}. The binding energy is found to be 173 $\pm$ 5 meV. Toluene has a factor of $\sim$ 30 increase in annihilation rate as compared to benzene, and is comparable to enhancements seen in alkane molecules \cite{Danielson23}.  

Shown in Fig.~\ref{fig:wide1}\,(c) is the annihilation spectrum of acetophenone. It is a benzene ring with an acetyl group (CH$_3$CO) attached, making it the simplest aromatic ketone. It has a net dipole moment of about $\sim 3$ D and four $\pi$ bonds.  Using the five phenyl and three methyl IR active vibration modes \cite{gambi1980infrared}, the binding energy is 288 $\pm$ 5 meV. 

Aniline is an aromatic compound comprised of a benzene ring attached to an amide group (-$\ce{NH2}$). It has a net dipole moment of about 1.5 D and three $\pi$ bonds. Figure \ref{fig:wide1}\,(d) shows the annihilation spectrum of aniline as a function of mean parallel energy. There is an enhanced resonant peak around 0.2 eV which is due to the phenyl C-H modes.   Using these modes, the resulting binding energy of aniline is 233 $\pm$ 5 meV. 
In comparison with toluene, both molecules show similar levels of enhancement of $Z_{\text{eff}}$ (i.e., to within an order of magnitude). The difference in $\varepsilon_{b}$ of about 57 meV can likely be associated with the larger dipole moment in aniline. Similar to other molecules, aniline spectrum displays broad spectral weight below 0.1 eV. While the shoulder near 0.19 eV on the high-energy side of the main peak is likely associated with the N-H mode, the current energy resolution is not good enough to make a firm identification.

Phenylacetylene (PA) and benzonitrile were also studied. In the PA molecule, one hydrogen on benzene is replaced with an alkyne, C$\equiv$C-H that  includes a CC triple bond. In benzonitrile, one hydrogen is replaced with a cyanide (C$\equiv$N) group, which also includes a triple bond. Thus, both molecules have an additional two $\pi$ bonds from the triple bonds, in addition to the three from the phenyl group, for a total of five $\pi$ bonds. One difference, however, is that the net dipole of PA ($\mu$ = 0.66 D) is about six times smaller than that of benzonitrile ($\mu$ = 4.52 D). 

Figure \ref{fig:wide1}\,(e) shows the measured annihilation spectrum for phenylacetylene.  Using the phenyl C-H modes to define the resonant peak, $\varepsilon_{b}=$ 230 meV $\pm$ 5 meV. It is surprising that there seems to be no obvious spectral feature that can be associated with the C$\equiv$C-H group. This is similar to the lack of an enhanced spectral feature that can be associated with the N-H modes in aniline.

The annihilation spectrum of benzonitrile is shown in Figure \ref{fig:wide1}\,(f). Substitution of the highly polar C$\equiv$N group creates a net dipole moment of 4.5 D, which is the largest dipole moment of any molecule in the present study. Analysis using the phenyl C-H stretch modes yields $\varepsilon_{b}=$ 298 meV $\pm$ 5 meV. The enhanced $\varepsilon_{b}$ value relative to PA is likely associated with increased dipole moment. The peak $\Zeff$ in BA is a factor of 3 \emph{lower} than in PA.

\begin{figure}
\includegraphics[width=1\linewidth,keepaspectratio]{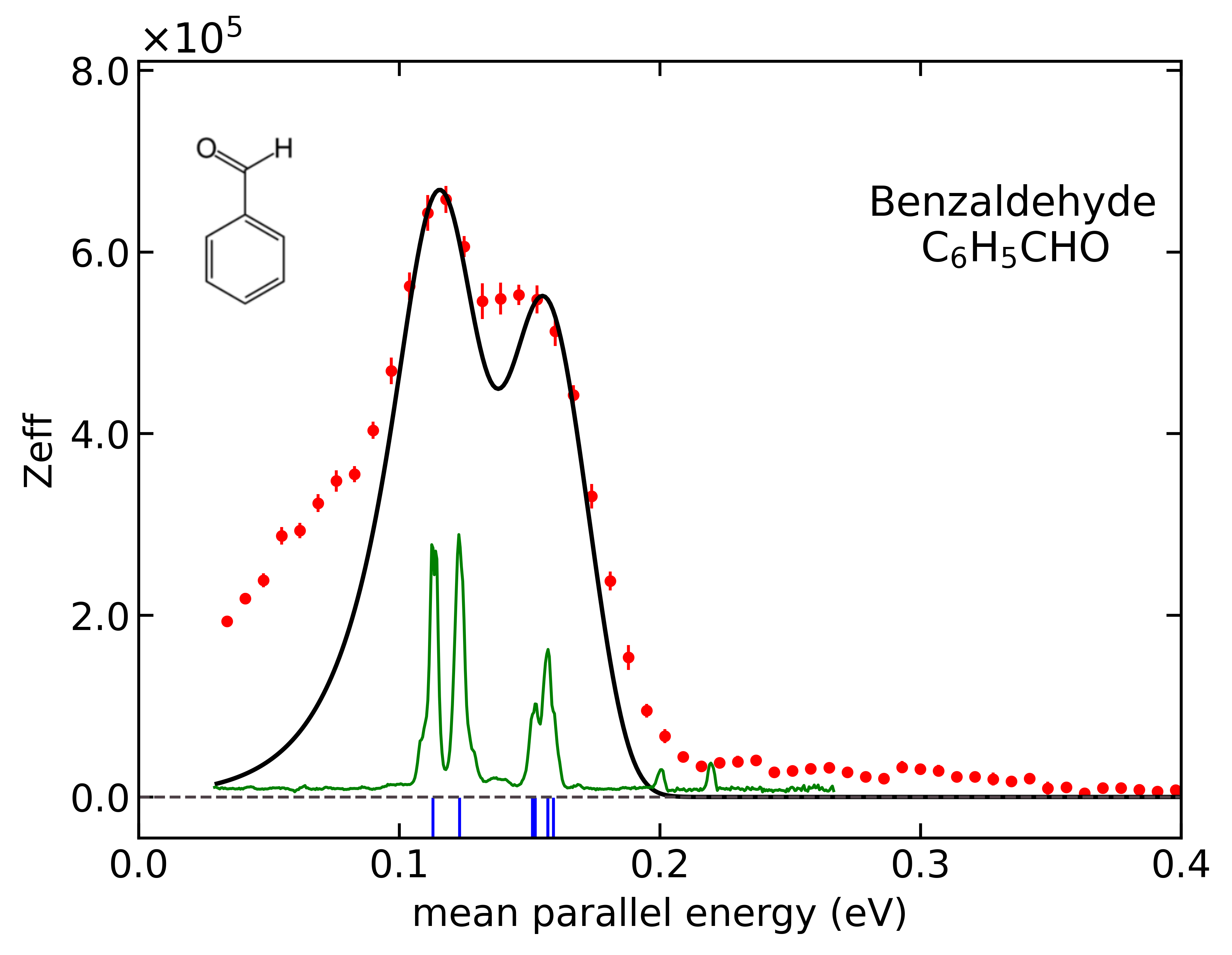}
\caption{\label{fig:benzaldehyde} $Z_{\text{eff}}$ measurements for benzaldehyde. The green curves are the IR spectra obtained from NIST \cite{linstorm1998nist}, and vertical blue bars indicate the locations of the IR active fundamental modes, with both downshifted by $\varepsilon_{b}$. The solid black curve is the fit (see text for details).}
\end{figure}

Also considered in this group is benzaldehyde (BA), which is benzene with an attached aldehyde (CHO) group, making it the simplest aromatic aldehyde. Substitution of the aldehyde group results in a relatively large dipole moment, $\mu$ = 3.14 D \cite{johnson2013nist,desyatnyk2005rotational}. The BA molecule has a total of four $\pi$ bonds, three from benzene and the fourth due to the C=O double (carbonyl) bond. The annihilation spectrum for BA is shown in Fig.~\ref{fig:benzaldehyde}. The spectrum shows two large peaks that are not completely resolved, one near 0.1 eV and one near 0.15 eV. The peak at 0.15 eV is due to the phenyl C-H stretch modes.

The lower-energy peak is in the region of the aldehyde C-H stretch mode. However, for BA, this mode is in a strong Fermi Resonance \cite{green1976vibrational} with the aldehyde C-H bend overtone. The result is two peaks with comparable amplitude, both of which are visible in the IR spectrum (cf.~Fig.~3). The energy resolution of the 300 K beam is not sufficient to determine whether the annihilation resonance is from one or both modes, but including this spectral feature is necessary to determine $\varepsilon_{b}$. 

The binding energy analysis used the vibrational frequencies of the two FR modes and the C-H vibrational modes from the phenyl group \cite{tolstorozhev2012ir,green1976vibrational}. The result is $\varepsilon_{b}$ = 220 $\pm$ 10 meV. The larger error bar arises from the uncertainty in identifying the correct mode frequencies for use with Eq.~2. This value of $\varepsilon_{b}$ for BA is a factor of $\sim1.7$ times larger than that for benzene ($\varepsilon_{b}$ = 132 meV; $\mu$ = 0 D) and $\sim1.2$ times larger than that for toluene ($\varepsilon_{b}$ = 173 meV; $\mu$ = 0.3 D). The difference likely reflects the larger dipole moment of BA. In contrast, BA is a factor of $\sim0.75$ times smaller than that of acetophenone ($\varepsilon_{b}$ = 288 meV; $\mu$ = 3.0 D), which has a comparable dipole moment but a larger $\alpha$ due to the added methyl group.

\subsection{Heterocyclic aromatics}
Four heterocyclic aromatic compounds were also studied: furan, pyrrole, pyridine and pyridazine. Heterocyclic molecules are compounds with at least two different atoms in the ring. This can change the molecular polarizability and electronic structure, and it often adds a significant dipole moment.

\begin{figure}
\includegraphics[width=1\linewidth,keepaspectratio]{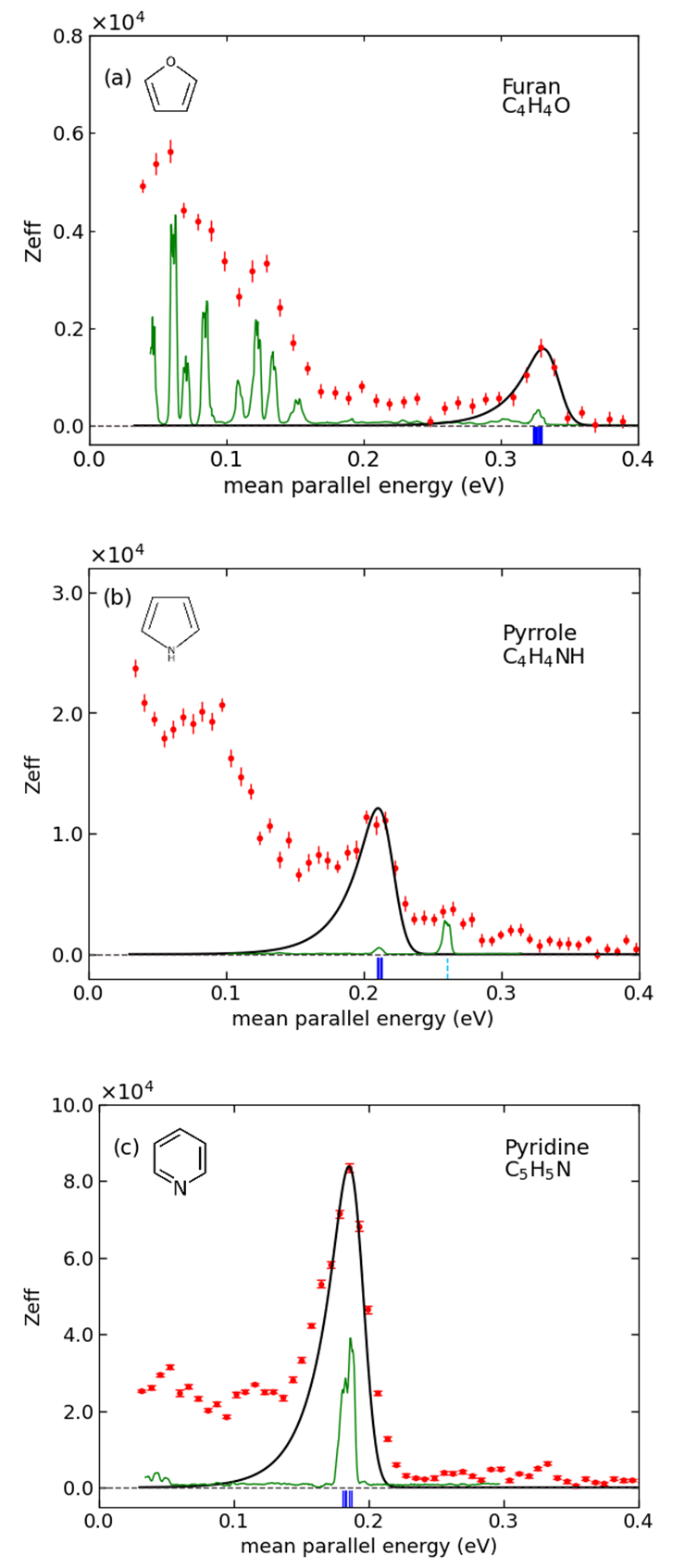}
\caption{\label{fig:widea} $Z_{\text{eff}}$ measurements for (a) furan, (b) pyrrole, and (c) pyridine. The green curves are the IR spectra obtained from NIST \cite{linstorm1998nist}, and vertical blue bars indicate the locations of the IR active fundamental modes, with both downshifted by $\varepsilon_{b}$. The solid black curves are fits to the C-H peaks. Vertical green dashed lines indicate dipole active modes not used in the analysis.}
\end{figure}

Furan, which is the smallest heterocyclic aromatic, is a five-member ring comprised of four carbon atoms and an oxygen. It has a dipole moment of 0.66 D. Figure \ref{fig:widea}\,(a) shows the $Z_\text{eff}$ spectrum. It exhibits a broad set of resonances below $0.15$ eV and a well resolved peak near $0.33$ eV. This later peak is identified as due to the phenyl C-H stretch modes \cite{mellouki2001vibrational}. Using Eq.~2, $\varepsilon_{b}= $ 52 $\pm$ 5 meV. From comparison to the downshifted IR spectrum, the unresolved lower-energy resonances correspond to a large number of low-energy fundamental modes, but were not analyzed in detail here. Such a spectrum is common for molecules with small $\varepsilon_{b}$ values where the low-energy vibrational mode resonances are prominent \cite{Jones12}. 

Figure \ref{fig:widea}\,(b) shows the annihilation spectrum of pyrrole. Pyrrole is a five-member ring similar to furan but with the oxygen atom replaced by an N-H group. Similar to furan, pyrrole has two $\pi$ bonds, but it has a larger dipole moment of $1.77$ D. The most distinct peak is near $0.2$ eV which is identified with the four phenyl C-H modes. The result is $\varepsilon_{b}$ = 165 $\pm$ 10 meV.  Similar to benzene, pyrrole has significant annihilation over a broad energy range where there are no fundamental modes. This leads to a larger uncertainty in the fit, and hence a larger error bar for $\varepsilon_{b}$. There is a small structure in the annihilation spectrum near $0.26$ eV that is likely associated with the single N-H vibration. This is consistent with the measured $\varepsilon_{b}$ but was not included in the analysis. As shown previously for benzene \cite{ghosh2022resonant}, the higher-resolution cryobeam will be necessary to resolve these features and obtain a more precise measurement of $\varepsilon_{b}$. 

Shown in Figure \ref{fig:widea}\,(c) is the measured annihilation spectrum for pyridine. Pyridine has a benzene-like structure with a C-H group replaced by a nitrogen. Similar to benzene, it has three $\pi$ bonds. Due to the substitution, it has a dipole moment of 2.19 D. The spectrum has a single prominent peak near $0.19$ eV which is associated with the phenyl C-H modes. The result is $\varepsilon_{b} = $ 186 $\pm$ 5 meV.  Although the peak is well resolved, similar to that in benzene, there is a broad, relatively flat, region of enhanced annihilation at energies below the C-H peak extending to the low-energy limit of the measurements. This is likely due to multimode VFRs. Surprisingly, the IR spectrum does not show significant IR activity over much of this region. Understanding these features is a topic of current research.

\begin{figure}
\includegraphics[width=0.9\linewidth,keepaspectratio]{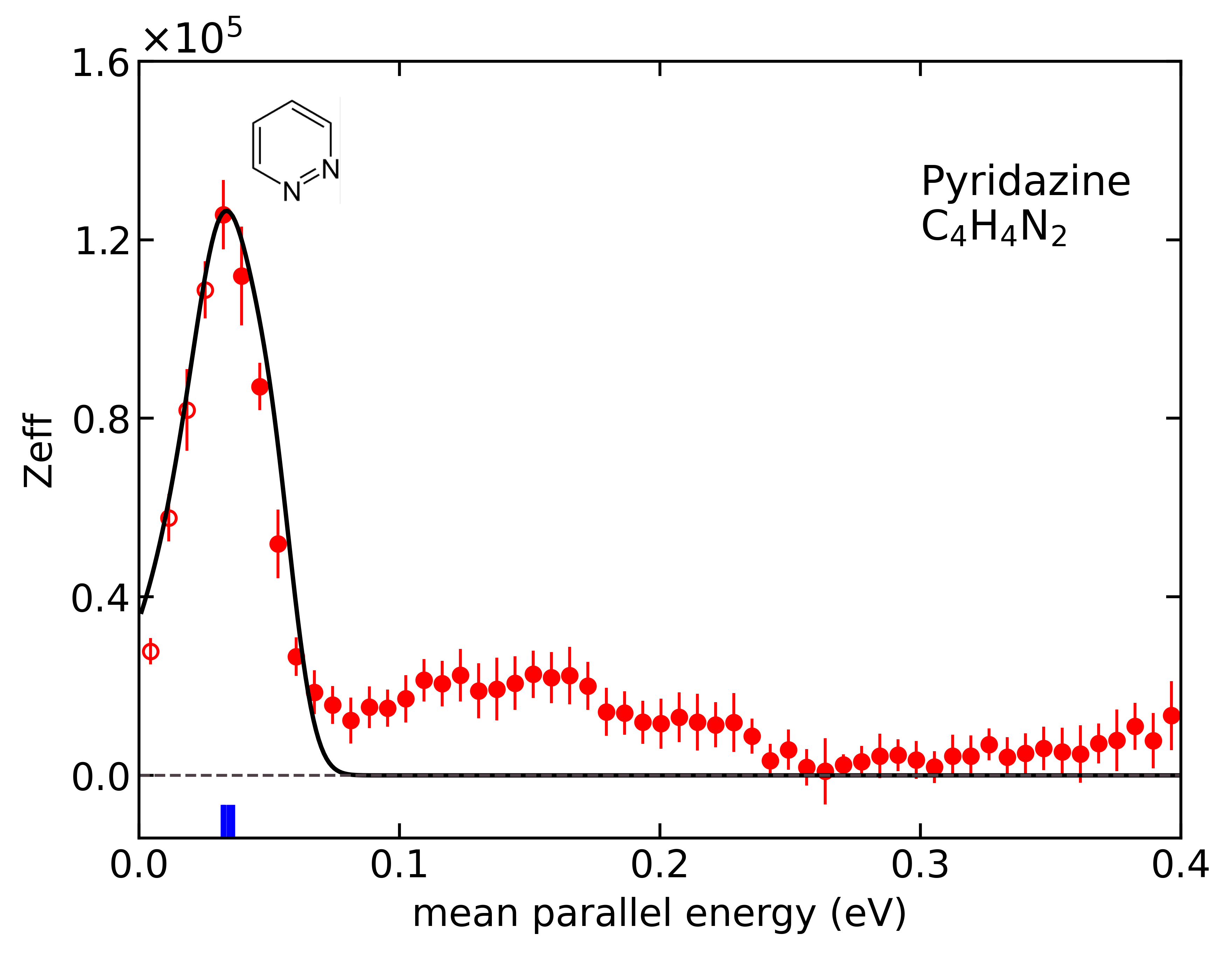}
\caption{\label{fig:pyr} $Z_{\text{eff}}$ measurements for pyridazine. The vertical blue bars indicate the locations of the IR active fundamental modes, with both downshifted by $\varepsilon_{b}$. The solid black curves are fits to the C-H peaks. Hollow circles are data where systematic errors cannot be ruled out (see text for details).}
\end{figure}

Pyridazine adds a second nitrogen to the pyridine structure and thus is a diazine. In this case, the dipole moment almost doubles to $4.2$ D, and so it is expected to have a significantly higher $\varepsilon_{b}$ value than pyridine. The measured annihilation spectrum is shown in Fig.~\ref{fig:pyr}. The spectrum shows an apparent peak near $0.04$ eV, which is at the low-energy limit of our measurements. In order to resolve the peak position, the data must be extended down to near $E_{\parallel} \approx 0$. Due to non-ideal effects, including approaching the beam cutoff and scattering effects at low energies, both the magnitude of the peak and position are not assured to be reliable. Neglecting this consideration, and ascribing the peak to the 4 phenyl CH mode yields $\varepsilon_{b}$ = 330 meV. Assuming the mode identification is correct, we can use the fit to the high-energy side of the peak to state that $\varepsilon_{b} \geq$ 330 meV (i.e., a lower bound on $\varepsilon_{b}$). Above the main peak, there is a broad range of enhanced annihilation extending to $\sim 0.25$ eV. This feature is similar to that seen in benzene and several deuterated benzenes \cite{ghosh2022resonant}. It is also prominent in many other aromatics [e.g., phenylacetylene (Fig.~2\,(e))], and is an important topic for future study.

\subsection{Non-aromatic cyclic molecules}

All of the molecules discussed above were aromatics. This raises the question as to whether, or how, aromaticity effects $\varepsilon_{b}$. In a previous study, we measured $\EB$ for several 5- and 6-carbon alkane rings at different levels of saturation \cite{Danielson23}, all of which had zero or very small dipole moments ($\mu < 0.5$). Those molecules exhibited modest increases of $\varepsilon_{b}$ with increasing $\Npi$, but they each had only one or two $\pi$ bonds. In order to extend the comparison to non-aromatic molecules with larger $\Npi$, we measured the spectra for cycloheptatriene (\ce{C7H8}, 3 $\pi$ bonds), and cyclooctatetraene (\ce{C8H8}, 4 $\pi$ bonds). The results and comparison to the analogous saturated 7 and 8 carbon alkanes are included in Table~1.

\begin{figure*}[ht]
\includegraphics[width=0.9\linewidth,keepaspectratio]{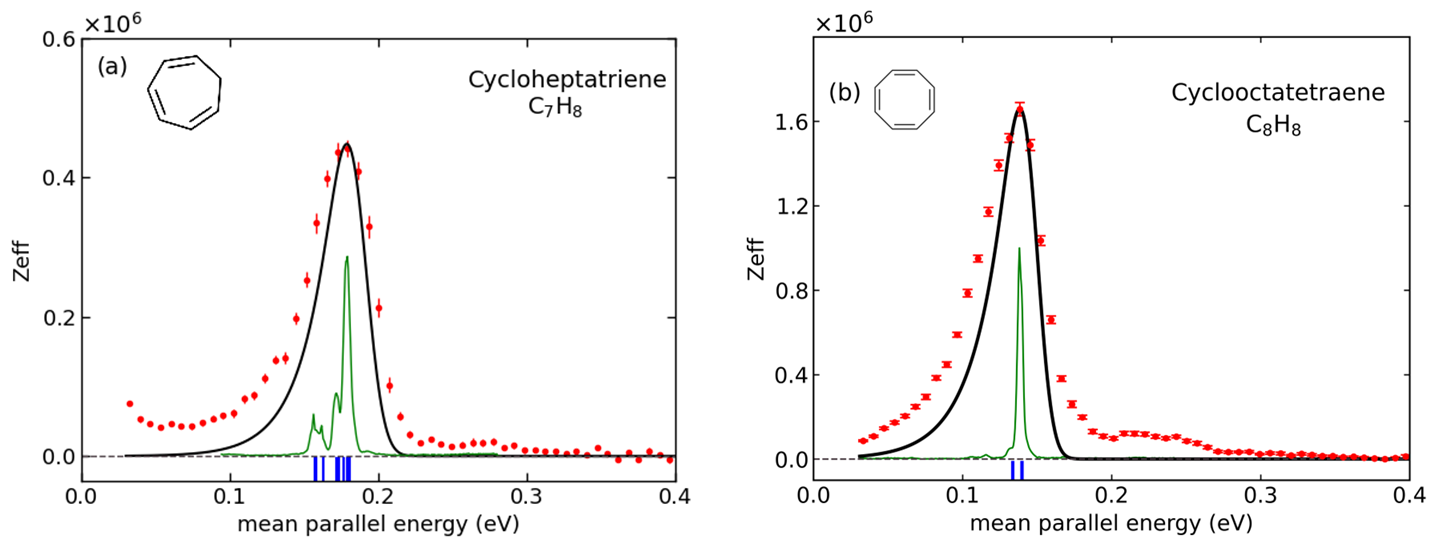}
\caption{\label{fig:wideb} $Z_{\text{eff}}$ measurements for the non-aromatic and ring molecules (a) cycloheptatriene and (b) cyclooctatetraene.  The green curves are the IR spectra obtained from NIST \cite{linstorm1998nist}, and vertical blue bars indicate the locations of the IR active fundamental modes, with both downshifted by $\varepsilon_{b}$. The solid black curves are fits to the C-H peaks.}
\end{figure*}

Spectra for cycloheptatriene (CHT) and cyclooctatetraene (COT) are shown in Figs.~$\ref{fig:wideb}$(a) and (b). In the case of CHT, a VFR peak is observed at $\simeq$ 0.17 eV, dominated by the C-H stretch vibrational modes. Here, the peak is broadened because there are two types of C-H modes with different characteristic energies. Using all 8 modes, we obtain $\varepsilon_{b} = $ 190 $ \pm 8$ meV. The curve fits reasonably well, but there appears to be extra spectral weight that is not accounted for on both sides of the peak. Further, the model assumed equal amplitude for all 8 modes. If that is incorrect, it could shift the measured $\varepsilon_{b}$ value by several milli electron volts leading to a larger uncertainty for this molecule.

For COT, there are 6 distinct C-H modes (2 degenerate) with 3 non-dipole active. Using only the dipole-active modes, $\varepsilon_{b}$ = 225 $\pm$ 5 meV. There also appears to be some broadening on the low energy side of the peak, but in contrast to CHT, the peak is well fit. Thus, this broadening does not appear to significantly affect $\varepsilon_{b}$. For both molecules (more so COT), there is a broad region of enhanced annihilation above the CH resonance. This is very similar to what was seen in the aromatics discussed above. While not presently understood, it appears that this extra structure is most prominent in molecules with appreciable symmetry.

With regard to the $\varepsilon_{b}$ values, although COT has no dipole moment, the presence of the extra $\pi$ bond and more symmetric structure of the molecule leads to a larger binding energy (i.e., a difference of $>$ 35 meV) compared with CHT. We can also compare the binding energies of these molecules with the saturated rings cycloheptane and cyclooctane  Ref.~\cite{Danielson23}. For CHT as compared to cycloheptane, adding 3 $\pi$ bonds increases $\varepsilon_{b}$ by $\sim 80$ meV.  For COT compared to cyclooctane, a larger increase of $\sim$ 100 meV is observed by adding the 4 $\pi$ bonds. This trend extends the results observed for the smaller rings \cite{Danielson23} to larger numbers of $\pi$ bonds and shows the strong impact that $N_{\pi}$ has on $\varepsilon_{b}$.

\section{\label{sec:Theory}THEORY OF POSITRON BINDING AND COMPARISONS WITH EXPERIMENTAL MEASUREMENTS}
We use the \emph{ab initio} many-body theory \cite{Hofierka22} to calculate positron binding energies for the molecules considered, assuming fixed nuclei \footnote{Note that vibrational and geometry relaxation effects are known to typically provide a few percent correction to fixed-nuclei calculations of binding energies and wave function densities \cite{APMO2014, Tachikawa14,Gianturco:posvibs,Buenker1,Buenker2,Grib10}.}. 
The approach is implemented in our {\tt EXCITON+} code \cite{Hofierka22}, which is heavily adapted from the original {\tt EXCITON} all-electron molecular many-body code \cite{Patterson2019} to include positrons. 
We solve the Dyson equation \cite{fetterwalecka,mbtexposed}
\begin{equation}\label{dyson_equation}
   \left( H^{(0)} + \hat{\Sigma}_{\varepsilon}\right) \psi_{\varepsilon}(\bf{r}) = \varepsilon \psi_{\varepsilon}(\bf{r}),
\end{equation}
where $\psi_{\varepsilon}(\bf{r})$ is the quasiparticle positron wave function with energy $\varepsilon$, $H^{(0)}$ is the Hamiltonian of the positron in the static (Hartree-Fock) field of the ground-state molecule and $\hat{\Sigma}_{\varepsilon}$ is the energy-dependent correlation potential (self energy of the positron in the field of the molecule) that accounts for electron-positron correlations. The energy dependence of $\Sigma$ demands a self-consistent solution of Eqn.~(\ref{dyson_equation}) for the bound state with $\varepsilon_b=|\varepsilon|$ for negative eigenvalues $\varepsilon$. Our approach includes three main diagrammatic contributions to the positron-molecule self energy (see Fig.~1 of \cite{Hofierka22}). The first is the $GW$ contribution, $\Sigma^{GW}$, which accounts for polarization of the molecular electron cloud by the positron, screening of the electron-positron Coulomb interaction (the random phase approximation ring series), and electron-hole attraction (time-dependent-Hartree-Fock or Bethe-Salpeter-Equation approximation, depending on whether bare or screened electron-hole Coulomb interactions are used within the ring series). The second contribution describes the nonperturbative process of virtual-positronium formation (where a molecular electron temporarily tunnels to the positron, leading to an attractive interaction), via calculation of the infinite ladder series of electron-positron interactions $\Sigma^{\Gamma}$. The third contribution is an analogous diagram containing the ladder series for positron-hole repulsion, $\Sigma^{\Lambda}$. We calculate the total self energy via their sum as: $\Sigma = \Sigma^{GW} + \Sigma^{\Gamma} + \Sigma^{\Lambda}$.

The positron and electron wave functions are expanded in distinct Gaussian basis sets, centred at each atom in the molecule and at five to ten additional `ghost' centres approximately 1$\mathring{\rm{A}}$ from the molecule. This combination enables a good description of the positron wave function at the nuclei and in the regions away from but close to the molecule where virtual-positronium formation occurs (see \cite{Hofierka22} for more details). For both the electron and positron bases, we use diffuse-function-augmented correlation-consistent polarized aug-cc-pVXZ (X=T, Q) Dunning basis sets \cite{Dunning}. An additional large even-tempered basis set is included for the positron at the region of highest positron density (for polar molecules), or at the center of the molecule (for non polars), to account for long-range correlations. The even-tempered bases have exponents $\zeta_k = 
\zeta_0\beta^{k-1}$, $k=1, 2,\ldots$ and take the form $10s\,9p\,8d\,7\!f\,3g$ with $\beta=2.2$, $\zeta_0=10^{-3}$ for all molecules except COT, benzene and PA, which have $10s\,9p\,8d\,7\!f\,6g$ with $\beta=2.0$ and $\zeta_0=10^{-3}$ (benzene and PA) or $\zeta_0=10^{-4}$ (COT). 
Solution of the Dyson equation for the molecules considered required diagonalization of Casida matrices of size ~0.5M$\times$0.5M, requiring $\sim$15\,TB RAM. They were performed on the UK Tier-2 HPC cluster Kelvin2, and the UK National Supercomputer ARCHER2.

The calculated binding energies are presented in Table \ref{tab:bind} at three successively more sophisticated levels of many-body theory which differ in the approximation used to calculate the ladder series for the virtual positronium formation and positron-hole repulsion self-energy contributions: the first (second) uses bare (screened) Coulomb interactions in the ladder diagrams and Hartree-Fock molecular orbital energies, and the third, our most sophisticated approach, uses screened Coulomb interactions and molecular orbital energies calculated at $GW$ level \cite{Patterson2019}. Additionally, results are shown for a model approach \cite{Hofierka22}, which approximates the computationally-expensive virtual-positronium formation self energy, $\Sigma^{\Gamma}$, by scaling the bare polarization self energy, $\Sigma^{(2)}$, by a factor $g$ so that $\Sigma \approx g\Sigma^{(2)} + \Sigma^{\Lambda}$. In the molecules studied to date \cite{Hofierka22, cassidy2023manybody} and here, setting $g=1.4$ and $g=1.5$ provides good lower and upper bounds for the positron binding energy, respectively.
Table \ref{tab:bind} also contains calculated values of the dipole moment $\mu$, dipole polarizability $\alpha$ and ionization energy $I$ which are in good agreement with the reference values in Table \ref{tab:table3}. 
The ionization energies are calculated at the HF and $GW$ levels. 
We typically find that it is only the energy of the highest occupied molecular orbital (HOMO) that is larger, and that the MOs below the HOMO have smaller energies at the $GW$ level compared to HF; as such the latter are more easily perturbed by the positron leading to an increase in the strength of the positron-molecule correlation potential, and ultimately increased binding energies compared to the calculations that use the HF energies in the diagrams.

\begin{table*}
	\caption{\label{tab:bind}
		Calculated MBT positron binding energies $\varepsilon_b$ (meV) for ringed molecules compared with experiment. 
		Also shown are dipole moments $\mu$ (calculated at HF), dipole polarizabilities $\alpha$ (calculated at the $GW$@BSE level of MBT), ionization energies $I$  (calculated at HF and the $GW$@RPA level of MBT) and annihilation contact densities (calculated at the $GW+\Gamma+\Lambda$ level of MBT with enhancement factors \cite{Green2015}). }
	\begin{ruledtabular}
		\begin{tabular}{l@{\hskip8pt}l@{\hskip12pt}c@{\hskip8pt} c@{\hskip8pt}c@{\hskip16pt} c@{\hskip8 pt}c@{\hskip8pt} c@{\hskip8pt} c@{\hskip8pt} c@{\hskip8pt}}
			&&\multicolumn{3}{c}{Molecular properties}&\multicolumn{3}{c}{Positron binding energy [meV]} &\\			\cline{3-5}
			\cline{6-8}
			\\[-1.5ex]
			Molecule& Formula &$\mu$\,[D] &$\alpha$\,[\AA$^3$] &$I$\,[eV]\footnote{The first (second) number is that calculated at HF ($GW$@RPA).}	
			& MBT\footnote{Many-body calculations at three levels of $\Sigma^{GW+\Gamma+\Lambda}$: (i) using bare Coulomb interactions within the ladders and HF energies; (ii) using dressed Coulomb interactions within the ladders and HF energies;  and (iii) using dressed Coulomb interactions and $GW$@RPA energies. The latter (highlighted in bold) is the most sophisticated calculation.}  	& Exp.   & {MBT model}\footnote{
Using a scaled self energy to account for virtual positronium formation \cite{Hofierka22}, $\Sigma\approx g\Sigma^{(2)}+\Sigma^{\Lambda}$, with $g=1.4$ and $g=1.5$.
} &$\delta$ [a.u.]\footnote{Figures in square brackets indicate powers of 10.}\\ 
& \\
\hline\\[-1ex]
{Benzonitrile}  & \ce{C6H5CN}  & 4.9 & 12.1 & 9.86, 10.07 & 326, 303, --\footnote{Our screened ladder with $GW$ energies calculation for benzonitrile is forthcoming; as such, in the text we compare theory and experiment only for the other 10 molecules in the table.}& 298$\pm$5 & 284, 343 & 1.60[$-$2] \\
{Pyridazine}  & \ce{C4H4N2}  & 4.3 & 8.2 & 10.57, 10.87 & 334, 308, {\bf 333} & 330$\pm$5 & 327, 398 & 1.89[$-$2] \\
{Benzaldehyde} 		& \ce{C6H5CHO}	& 3.5 & 11.9 & 9.64, 9.90 & 210, 192, {\bf 213} & 220$\pm$10 & 198, 256 &1.46[$-2$] \\
{Pyridine} 			& \ce{C5H5N}		& 2.3 & 8.9 & 9.52, 9.91 & 184, 155, {\bf 181} & 186$\pm$5 & 149, 211 &  1.58[$-2$] \\
{Pyrrole} 			& \ce{C4H4NH} 		& 1.9 & 7.6  & 8.22, 8.64  & 157, 127, {\bf 148} & 165$\pm$10 & 94, 148  &1.55[$-$2] \\
{Aniline} 			& \ce{C6H5NH2} 		& 1.4 & 11.0 & 8.07, 8.26  & 220, 180, {\bf 209} & 233$\pm$5 & 172, 251  & 1.92[$-$2] \\
{Phenylacetylene} 	& \ce{C6H5CCH}		& 0.8 & 13.3 &8.86, 9.15 & 231, 187, \textbf{220} & 230$\pm$5 & 183, 266 & 1.87[$-2$] \\
{Furan} 			& \ce{C4H4O} 		& 0.7 & 6.7 & 8.85, 9.35 & 45, 32, {\bf 42} & 52$\pm$5 & 20, 49 & 7.32[$-$3]\\
{Toluene} 			& \ce{C6H5CH3} 	& 0.4 & 11.4 & 8.87, 9.14 & 174, 135, {\bf 161} & 173$\pm$5 & 128, 203  & 1.87[$-2$]\\ 
{Benzene} 			& \ce{C6H6} 	& 0.0 & 9.8 & 9.21, 9.57 & 158, 122 {\bf 148} & 132$\pm$3&  &  1.61[$-2$] \\ 
{Cyclooctatetraene} 	& \ce{C8H8}\footnote{In the `tub' ($D_{2\rm d}$) geometry.}		& 0.0 & 13.0 & 8.37, 8.66 & 202, 158, {\bf 190} & 225$\pm$5 & 151, 238 
& 2.03[$-2$] \\[1ex]
\hline
\multicolumn{9}{l}{\includegraphics*[width=0.94\textwidth]{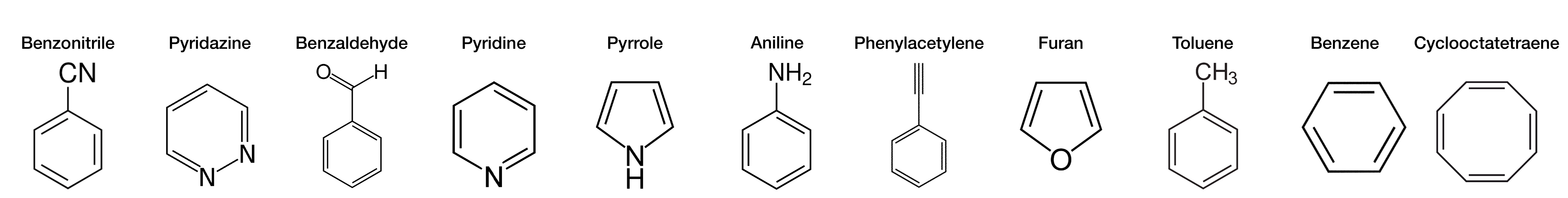}}
		\end{tabular}
	\end{ruledtabular}
\end{table*}

Comparison of the predicted and measured binding energies is shown in Fig.~7. Agreement is good to excellent. For ten of the molecules for which $\varepsilon_{b}$ is given in Table \ref{tab:bind}, the rms difference in binding energy between theory (using values in bold) and experiment is 13 meV. This corresponds to an rms fractional deviation in the experimental and calculated binding energies,  $[\varepsilon_{b}$(calc.) $- \varepsilon_{b}$(exp.)]/$\varepsilon_{b}$(exp.) of 9\%. 
One noticeable trend is that, of the comparisons presented here, the measured binding energies of six of the ten molecules are larger than the theoretical prediction, three are within the experimental error bars, and only one measured value (benzene) is smaller than that predicted \footnotemark[1].

Finally, having the positron bound-state wave function enables calculation of the electron-positron annihilation contact density $\delta$ 
as 
\begin{equation}\label{eqn:cd}
    \delta = \sum_{n=1}^{N_{\rm e}} \gamma_n \int \left| \phi_n(\mathbf{r})\right|^2\left| \psi_{\varepsilon}(\mathbf{r})\right|^2 d\mathbf{r},
\end{equation}
where $\phi_n(\mathbf{r})$ are the occupied electronic molecular orbitals, the sum is over the $N_{\rm e}$ occupied electron orbitals and $\gamma_n$ are enhancement factors which account for short-range electron-positron attraction, and depend on the $GW$ ionization energy $\varepsilon_n$ of each molecular orbital as follows \cite{Green2015}: $\gamma_n = 1 + \sqrt{1.31/\left|\varepsilon_n\right|} + (0.834 / \left|\varepsilon_n\right|)^{2.15}$. 

\begin{figure}[t]
\includegraphics[width=\linewidth,keepaspectratio,trim={0.5cm 0 0.5cm 1cm}]{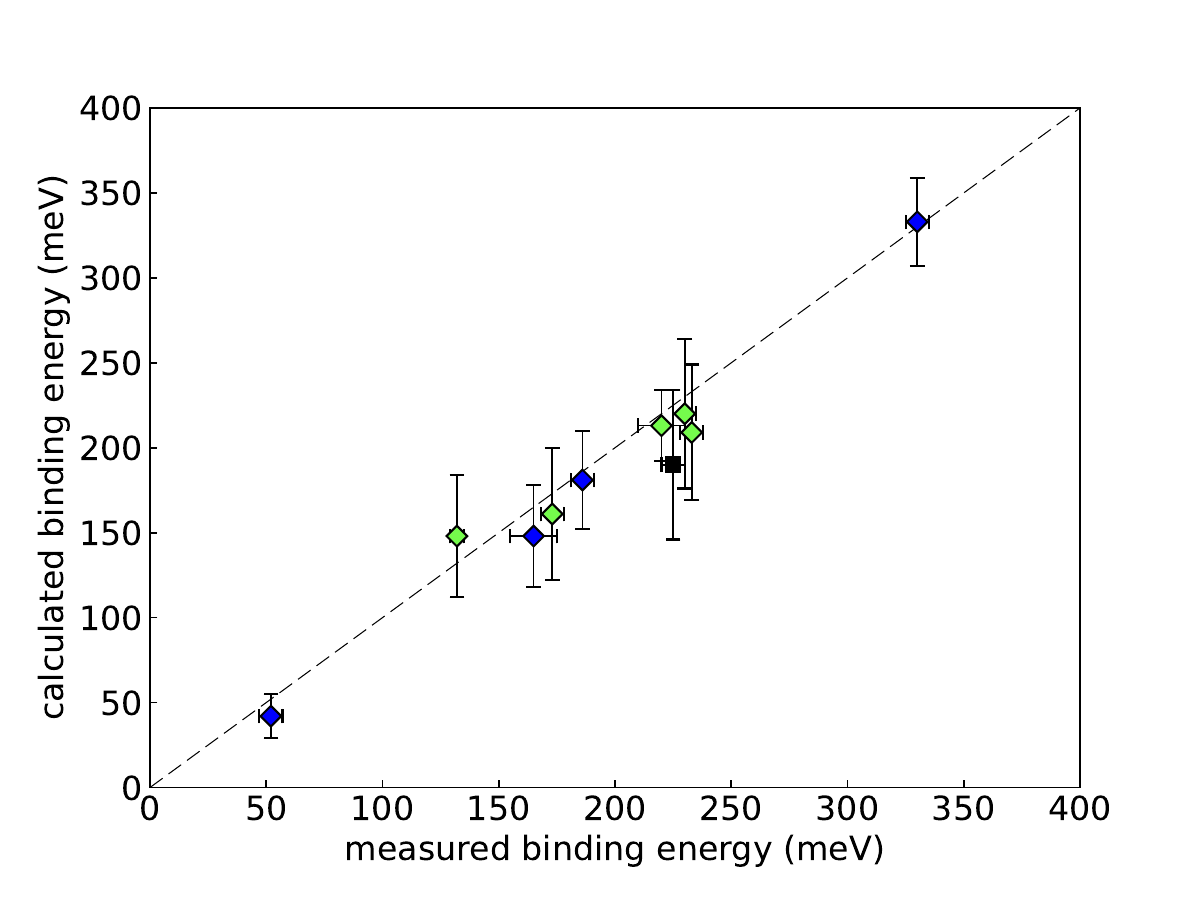}
\caption{\label{fig:wide2} Comparison of the calculated binding energies versus the measured binding energies. Green diamonds: benzene and benzene substitutions; blue diamonds: heterocyclic aromatic molecules; black square: COT. 
Error bars on the calculated values are the maximum difference between results from the three many-body calculations in Table \ref{tab:bind}.}
\end{figure}

\section{\label{sec:level1}FURTHER DISCUSSION}

\subsubsection{Localization of the bound-state positron wave function}
 Early \textit{ab initio} and model-potential calculations predicted $\varepsilon_{b}$ values and enabled the investigation of the shape of the positron wave function \cite{tachikawa2011ionization,tachikawa2012bound,Swann19,swann2020effect}.  When compared to the experimental measurements for $\varepsilon_{b}$, these calculations yielded considerable insight into roles of $\alpha$, $\mu$, and the molecular geometry for specific molecular series including saturated chain and ring alkanes \cite{Swann19,Danielson23} and simple halogen substitutions \cite{Swann21,suzuki2020machine}.  However, these theories were unable to adequately treat electron-positron correlations, including the effects of virtual positronium formation, and thus they were unable to achieve quantitative agreement with the experimental measurements.  

While the model-potential calculations have been effective in achieving agreement with the experimental measurements, they required fit parameters; and thus they have not been very successful in making predictions beyond the  molecular series for which the parameters were tuned \cite{Swann19,Swann21,Tachikawa14,Tachikawa20,Sugiura19}.  Further, most of the previous molecules studied that included $\pi$ bonds also had large dipole moments. This strongly localized the bound positron and dominated the binding energy (e.g., acetonitrile), and hence it tended to mask the influence of the $\pi$ bonds. Generally, the previous calculations either showed a more-or-less uniform positron cloud around the molecular periphery (e.g. alkanes), or a strongly localized cloud near the negative end of the molecule (e.g. acetonitrile), although other shape dependences have also been noted \cite{Swann21,danielson2021influence}.

In contrast, the MBT calculations are in quantitative agreement with experimental measurements, and they provide further insight into the effects that contribute to positron binding \cite{Hofierka22,cassidy2023manybody}. Beyond the influence of $\alpha$ and $\mu$, the calculations investigated the possible dependence on other parameters such as $I$ and $\Npi$. The effect of $I$ on binding is predicted to be weak, since binding depends on positron interactions with many orbitals, not just the highest occupied orbital that sets $I$. Successively deeper lying molecular orbitals generally contribute less to the positron-molecule correlation potential, but the decrease is not monotonic due to the enhanced strength of $\pi$ bonds. 
The $\pi$-bond orbital electron density is localized away from the ionic cores, allowing these electrons interact strongly with the positron \cite{Danielson09,Hofierka22}. Thus $\Npi$ can be an important parameter in determining $\varepsilon_{b}$, especially if the $\pi$ bonds are HOMOs (as in COT).

\begin{figure*}
\includegraphics[width=1.0\linewidth, keepaspectratio]{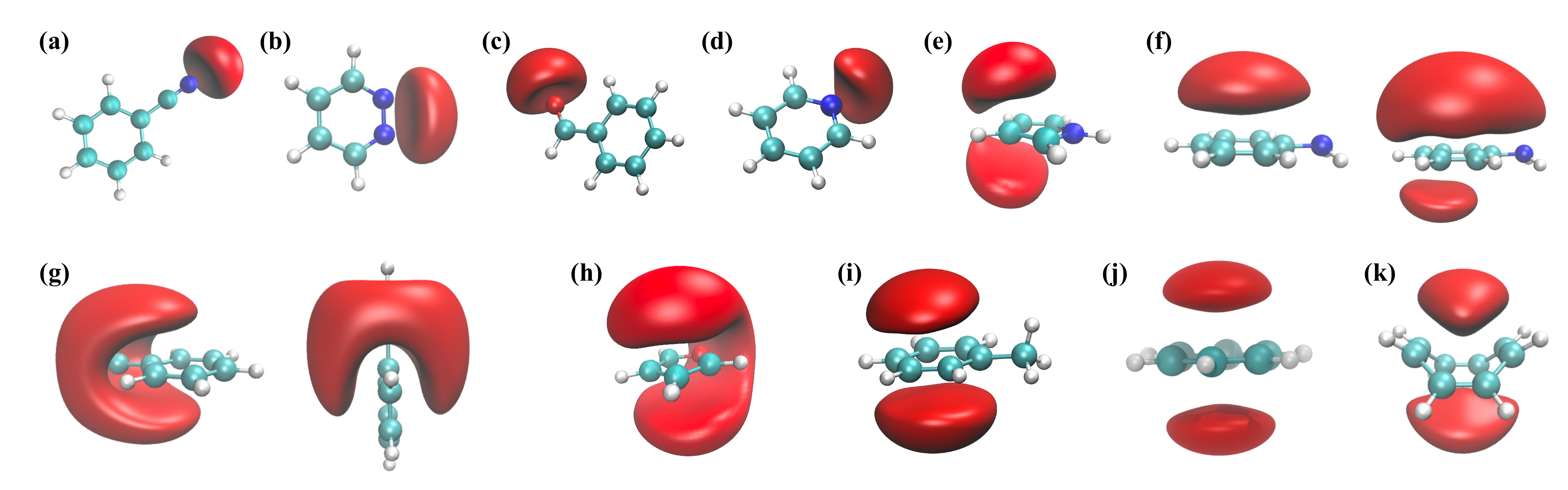}
\caption{\label{fig:boundstates} Correlated positron Dyson wave functions (positive and real as displayed here) for several heterocyclic molecules at the $GW+\Gamma+\Lambda$ level of many-body theory, shown at 80$\%$ of the wave function maximum unless otherwise stated. (a) benzonitrile; (b) pyridazine; (c) benzaldehyde; (d) pyridine; (e) pyrrole; (f) aniline (left, 80$\%$; right, 60$\%$); (g) phenylacetylene (two orientations, both 80$\%$); (h) furan; (i) toluene;  (j) benzene; (k) cyclooctatetraene ($D_\text{2d}$ geometry).}
\end{figure*}

The bound-state positron wave functions for the molecules studied here are shown in Fig.~\ref{fig:boundstates}. For molecules where the aromatic $\pi$ bonds dominate, the positron wave function is localized in planes above and below the plane of the molecule where the $\pi$-bond electron density is largest. Examples of this effect are seen in Figs. \ref{fig:boundstates}\,(j) and (i), benzene and toluene. In contrast, similar to what was seen with non-aromatic molecules \cite{Hofierka22}, the calculations show that the introduction of a large permanent dipole moment results in localization of the wave function very near the dipole, as seen in Figs. \ref{fig:boundstates} (c), (b) and (d) for benzaldehyde, pyridazine and pyridine.

However, the effects of attraction of the positron to the $\pi$-bond electrons and to a permanent dipole compete when both are present on the same molecule. As examples, aromatic $\pi$-bond localization is seen to be (weakly) perturbed by the dipoles in Figs. \ref{fig:boundstates}\,(f) and (e), aniline and pyrrole. A stronger effect is seen in (h), furan, where the 80\% positron contour extends from the region of the $\pi$-bond electrons all the way to the dipole. The C$\equiv$C triple-bond $\pi$ orbitals in (g) phenylacetylene result in concentration of the positron around this bond, which extends continuously into the region of the aromatic $\pi$-bond electrons. Lower-density contours would show larger regions of both the $\pi$-bond and triple bond localization.

Changes in molecular symmetry can also affect $\pi$-bond localization. An example of this is the distorted positron wave function contour resulting from the broken planar symmetry of 
cyclooctatetraene. Figure \ref{fig:boundstates}\,(f) for aniline shows that the strong, slightly-out-of-plane dipole results in restricting the 80\% positron wave function contour to the region above the plane, whereas the 65\% contour shows some amplitude below the plane as well. For all molecules, the long-range $\alpha$ potential results in a lower-amplitude, more uniform cloud of positron density further from, and entirely surrounding the molecule. 

\subsubsection{Relationship of $\varepsilon_{b}$ to molecular parameters}
\begin{figure*}
\includegraphics[width=0.7\linewidth,keepaspectratio]{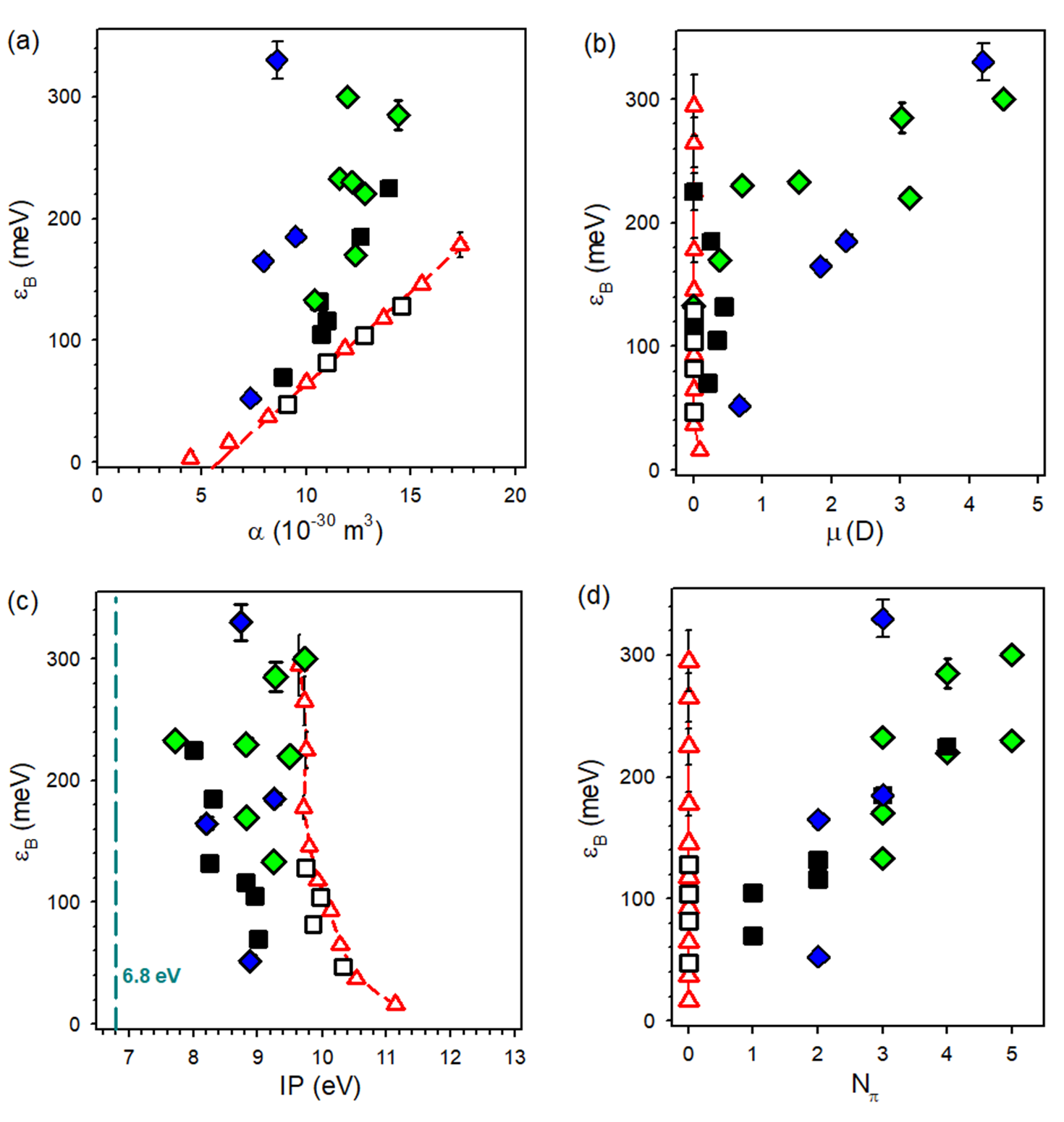}
\caption{\label{fig:molprop} Measurements of $\varepsilon_{\text{\textit{b}}}$  for aromatic and ring molecules vs. (a) molecular polarizability $\alpha$, (b) permanent dipole moment $\mu$, (c) ionization potential $I$, and (d) number of $\pi$ bonds N{$_\pi$}. Diamonds: six-member carbon aromatic rings: green, benzene and benzene substitutions; blue, heterocyclic molecules. Squares: five-member non-saturated and non-aromatic carbon rings.  Red triangles, alkanes. Black symbols, cyclic molecules. Solid symbols: one or more $\pi$ bonds and open symbols, no $\pi$ bonds. Data for alkanes from Ref. \cite{Danielson23} and other non-saturated rings from Ref. \cite{Dan22}.}
\end{figure*}

It has been natural to try to relate binding energies to global molecular parameters such as $\alpha$, $\mu$, $I$ and $N_{\pi}$, and much has been written on this topic. It has also been shown by comparison of isomers that molecular geometry can play an important role in positron binding, irrespective of other global parameters \cite{Dan21,swann2020effect}. Shown in Fig.~\ref{fig:molprop} are $\varepsilon_{b}$ values for the molecules studied here and selected previous data as a function of $\alpha$, $\mu$, $I$ and $N_{\pi}$, extending the ranges of parameters studied beyond those done previously.

There is naturally a large scatter in the data in the one-parameter-at-a-time representations in Fig.~\ref{fig:molprop}, since wide variations of the other parameters (not shown in a particular plot) produce changes in $\varepsilon_{b}$ as well. Despite the large scatter in the data, the largest changes of $\varepsilon_{b}$ are associated with changes in $\mu$ and $N_{\pi}$ [cf. panels (b) and (d)]. This highlights the role of $\mu$ and $N_{\pi}$ in determining $\varepsilon_{b}$ for the molecules studied here, similar to that seen in the spatial distribution of  the positron wave function.

From the data in Fig.~9 (c), $\varepsilon_{b}$ appears to be independent of $I$, to within the scatter in the values. This is made clearer when one considers separately $I$ for aromatic and non-aromatic rings. They span narrower bands of $I$ at differing mean $I$ (i.e., separated by $\sim$ 0.8 eV) but exhibit similar broad ranges of $\varepsilon_{b}$. This tends to confirm that $\varepsilon_{b}$ depends only weakly, if at all, on $I$ for the molecules studied here.

Regarding the dependence of $\varepsilon_{b}$  on $N_{\pi}$, the molecules shown in Fig.~9 include different types of $\pi$ bonds: aromatic and non-aromatic $\pi$ bonds in rings, $\pi$ orbitals in C=O double bonds, and $\pi$  bonds in the C$\equiv$C triple bond in phenylacetylene. When all are included [cf. Fig.~9\,(d)], they indicate that $\varepsilon_{b}$ increases with increasing $N_{\pi}$. While these limited data constrain detailed analysis, it also appears that the change in $\varepsilon_{b}$ with $N_{\pi}$ is similar for the aromatic and non-aromatic $\pi$ bonds. Thus, the present data are consistent with all $\pi$ bonds contributing approximately equally to $\varepsilon_{b}$, but more precise conclusions will require further study.

\subsubsection{Multimode phenomena in ring molecules}
A theory of coupling to dipole-allowed transitions is successful in explaining VFR due to fundamental modes in small molecules \cite{Grib06}, and there were two observations of similar quadrupole coupling \cite{Natisin17}. Going further, it is natural to try to associate annihilation VFR not associated with dipole- and quadrupole-allowed fundamental vibrations with other infrared activity in the molecule. However, a detailed explanation is lacking of many of the observed regions of enhanced annihilation that do not appear to correspond to fundamental vibrational modes. The working assumption is that these resonances are due to as yet unidentified multimode resonances (e.g., combinations and overtones) but this has yet to be confirmed. The special case of the Fermi resonance (FR) in benzaldehyde offers an opportunity to study such multimode resonances, for a case in which the individual modes have been identified. Better positron beam resolution will be needed to study this FR in further detail. 

There are also regions in the annihilation spectra where broad VFR activity is observed that do not appear to correspond to appreciable IR activity. Examples presented here are the spectra of benzene, pyridine and pyrrole. This raises the question as to whether there is some other, presently unrecognized, annihilation VFR mechanism (i.e., not visible in IR spectral measurements), but this is beyond the scope of the present work.

\section{\label{sec:level1}SUMMARY AND CONCLUDING REMARKS}

There has been much progress in understanding positron binding to molecules, see e.g., \cite{Grib06,Grib09,Grib10,Sugiura19,Tachikawa14,Tachikawa20,Swann19,Swann21,Suguira20, Hofierka22,cassidy2023manybody}, but many questions remain. This paper presents a study of resonant positron annihilation and positron binding in selected aromatic and ring compounds. It enabled an investigation of the effects of permanent dipole moment and $\pi$ orbitals, and the competition of the two, in determining positron annihilation and binding. Experimental measurements for $\varepsilon_{\textit{b}}$ were compared with the predictions of a recent \emph{ab initio} many body theory. Good-to-excellent agreement was found (i.e., 9\% rms fractional deviation between measurements and predictions for 10 molecules). This is was made more significant by the fact that the compounds span a wide range of {$N_\pi$} (2 - 5) and $\mu$ (0.3 - 4.5 D) and a factor of 1.7 in $\alpha$.

An important result of the theory is predictions for the spatial distribution of the bound positron wave function. This can be used to distinguish the different character of the bound states due to specific molecular parameters. As established  previously, the polarizability results in a more-or-less uniform bound state density surrounding the molecule, while a permanent dipole moment localizes the bound state density near the dipole. Aromatic $\pi$ bonds produce maxima in positron density above and below the planes of the ring molecules. A related $\pi$-bond localization effect is seen in the non-valence anion C$_{6}$F$_{6}^-$ in which the excess (weakly bound) electron has a very similar charge distribution to that of a positron attached to benzene, C$_6$H$_6$ \cite{voora2014nonvalence,rogers2018evidence}. 

This qualitative picture of positron wave function localization provides insight into positron-molecule bound states complementary to the quantitative predictions of the theory. An interesting question for future study is whether one could distinguish experimentally the $\pi$ and $\mu$ components of the bound state wave function using gamma-ray spectroscopy [i.e., Doppler-broadening or angular correlation of annihilation radiation (ACAR)] \cite{Iwat97}.

Understanding the observed enhancements in $Z_{\text{eff}}$ beyond the predictions of simple theory \cite{Grib06} is also a major gap in our understanding. One example of the chemical specificity discussed here include the increase in $Z_{\text{eff}}$ by a factor of 30 when a methyl group is added to benzene. Another example is the large amount of likely multimode spectral weight at low energies in pyrrole, relative to that due to the C-H stretch mode. That the observed annihilation VFR can be smaller, or orders of magnitude larger, than the predictions of simple theory has been attributed to vibrational coupling to other modes (i.e., IVR) \cite{Grib10}, however attempts to pursue this have had only modest success \cite{Grib17b}. Similarly, there is now appreciable evidence of multimode annihilation VFR (e.g., in benzene, benzaldehyde and furan), but little understanding of when and how this occurs. The observation of the annihilation Fermi resonance in benzaldehyde may provide further insight into this phenomenon.

\acknowledgements{
We wish to acknowledge helpful conversations with G. F. Gribakin and K. D. Jordan.
The work at UCSD was supported by the U.~S.~NSF, grant PHY-2306404 and the U.~C.~San Diego Foundation. The QUB theoretical work was funded by the European Research Council, grant 804383 `ANTI-ATOM' and used the Northern-Ireland HPC Service and the ARCHER2 UK National Supercomputing Service. 
}
\section*{Author Contributions}
J. P. Cassidy, S. K. Gregg and J. Hofierka contributed equally to the theoretical work.


\providecommand{\noopsort}[1]{}\providecommand{\singleletter}[1]{#1}%

\end{document}